\documentclass[reprint,amssymb, amsmath, aps, longbibliography, superscriptaddress, showpacs, footinbib, prl]{revtex4-1}
\usepackage[T1]{fontenc}
\usepackage[utf8]{inputenc}
\usepackage{amssymb}
\usepackage{amsmath}
\usepackage{color}
\usepackage{graphicx}
\usepackage{footmisc,booktabs}
\usepackage{gensymb}
\usepackage{sidecap}

\begin{document}

\title{Electronic signature of the vacancy ordering in NbO (Nb$_{3}$O$_{3}$)}

\author{A. K. Efimenko}
\affiliation{Max Planck Institute for Chemical Physics of Solids, Noethnitzer Str. 40, 01187 Dresden, Germany}

\author{N. Hollmann}
\affiliation{Max Planck Institute for Chemical Physics of Solids, Noethnitzer Str. 40, 01187 Dresden, Germany}

\author{K. Hoefer}
\affiliation{Max Planck Institute for Chemical Physics of Solids, Noethnitzer Str. 40, 01187 Dresden, Germany}

\author{J. Weinen}
\affiliation{Max Planck Institute for Chemical Physics of Solids, Noethnitzer Str. 40, 01187 Dresden, Germany}

\author{D. Takegami}
\affiliation{Max Planck Institute for Chemical Physics of Solids, Noethnitzer Str. 40, 01187 Dresden, Germany}

\author{K. K. Wolff}
\affiliation{Max Planck Institute for Chemical Physics of Solids, Noethnitzer Str. 40, 01187 Dresden, Germany}

\author{S. G. Altendorf}
\affiliation{Max Planck Institute for Chemical Physics of Solids, Noethnitzer Str. 40, 01187 Dresden, Germany}

\author{Z. Hu}
\affiliation{Max Planck Institute for Chemical Physics of Solids, Noethnitzer Str. 40, 01187 Dresden, Germany}

\author{A. D. Rata}
\affiliation{Max Planck Institute for Chemical Physics of Solids, Noethnitzer Str. 40, 01187 Dresden, Germany}

\author{A. C. Komarek}
\affiliation{Max Planck Institute for Chemical Physics of Solids, Noethnitzer Str. 40, 01187 Dresden, Germany}

\author{A. A. Nugroho}
\affiliation{Insitiut Teknologi Bandung, Jl. Ganesha 10, 40132 Bandung, Indonesia}

\author{Y. F. Liao}
\affiliation{National Synchrotron Radiation Research Center (NSRRC), 101 Hsin-Ann Road, 30077 Hsinchu, Taiwan}

\author{K.-D. Tsuei} 
\affiliation{National Synchrotron Radiation Research Center (NSRRC), 101 Hsin-Ann Road, 30077 Hsinchu, Taiwan}

\author{H. H. Hsieh}
\affiliation{Chung Cheng Institute of Technology, National Defense University, Taoyuan 335, Taiwan}

\author{H.-J. Lin}
\affiliation{National Synchrotron Radiation Research Center (NSRRC), 101 Hsin-Ann Road, 30077 Hsinchu, Taiwan}

\author{C. T. Chen}
\affiliation{National Synchrotron Radiation Research Center (NSRRC), 101 Hsin-Ann Road, 30077 Hsinchu, Taiwan}

\author{L. H. Tjeng}
\email{Hao.Tjeng@cpfs.mpg.de}
\affiliation{Max Planck Institute for Chemical Physics of Solids, Noethnitzer Str. 40, 01187 Dresden, Germany}

\author{D. Kasinathan}
\email{Deepa.Kasinathan@cpfs.mpg.de}
\affiliation{Max Planck Institute for Chemical Physics of Solids, Noethnitzer Str. 40, 01187 Dresden, Germany}

\date{\today}

\begin{abstract}
We investigated the electronic structure of the vacancy-ordered 4$d$-transition metal monoxide NbO (Nb$_3$O$_3$) 
using angle-integrated soft- and hard-x-ray photoelectron spectroscopy as well as ultra-violet angle-resolved 
photoelectron spectroscopy. We found that density-functional-based band structure calculations can describe the spectral 
features accurately provided that self-interaction effects are taken into account. In the angle-resolved spectra we 
were able to identify the so-called vacancy band that characterizes the ordering of the vacancies.  This together with 
the band structure results indicates the important role of the very large inter-Nb-$4d$ hybridization for the
formation of the ordered vacancies and the high thermal stability of the ordered structure of niobium monoxide. 
\end{abstract}


\maketitle

The transition metal monoxide NbO is special. While many transition metal monoxides adopt the highly
dense rocksalt crystal structure, NbO, synthesized already more than 150 years ago \cite{Rose_1858},
crystallizes in a structure in which 25\% of the Nb and 25\% of the O ions are removed from the
rocksalt lattice. The Nb and O vacancies are ordered and there are no additional distortions of the
lattice \cite{Brauer_1941,Andersson_1957,Bowman_1966}, see Fig.\,1. For clarity, the hypothetical rocksalt and 
the actual vacancy-ordered crystal structure of niobium monoxide will be referred to as Nb$_{4}$O$_{4}$ 
and Nb$_{3}$O$_{3}$, respectively. Adding to the astonishment, the vacancy order is robust and crystal 
structure remains stable up until 2213 K, the melting point of niobium monoxide
\cite{Andersson_1957,Kolchin_1961,Taylor_1971_1,Taylor_1971_2,Reed_2007}. 

While the formation of defects or vacancies in either the cation or anion sites is in itself not a rare phenomenon
for transition metal (TM) oxides \cite{Wadsley_1964, Schafer_1964, Burdett_1995}, the precise and robust 
ordering of the vacancies in niobium monoxide give perhaps credit to efforts that treat its crystal structure 
not so much as a defect problem but as a three-dimensional network of corner-sharing condensates of 
Nb$_6$O$_{12}$ clusters. This is motivated by the analogy with the TM$_{6}X_{12}$ clusters that form 
Chevrel phases ($X$ = halogen, chalcogen or pnictide) \cite{Schafer_1964,Andersen_1984,Chevrel_1986}. 
Fig.\,1\,(right) gives an illustration of such a building block.

Density functional based band structure calculations have been performed
\cite{Wimmer_1982,Schwarz_1984,Andersen_1984,Schulz_1993}
to explain the properties of niobium monoxide \cite{Okaz_1975,Honig_1976,Erbudak_1978,Schulz_1992} 
and it was found that the Nb$_{3}$O$_{3}$ structure is more stable than the Nb$_{4}$O$_{4}$ by 
about 1\,eV per NbO unit \cite{Schwarz_1984,Schulz_1993}. Our objective here is to test experimentally 
the accuracy of band structure calculations and to identify the bands in the measured spectra that are 
characteristic for Nb$_{3}$O$_{3}$, thereby proving that the vacancies are ordered and that band 
formation is an essential ingredient for the formation of vacancies in niobium monoxide.

\begin{figure}[t]
\includegraphics[clip,width=0.48\columnwidth]{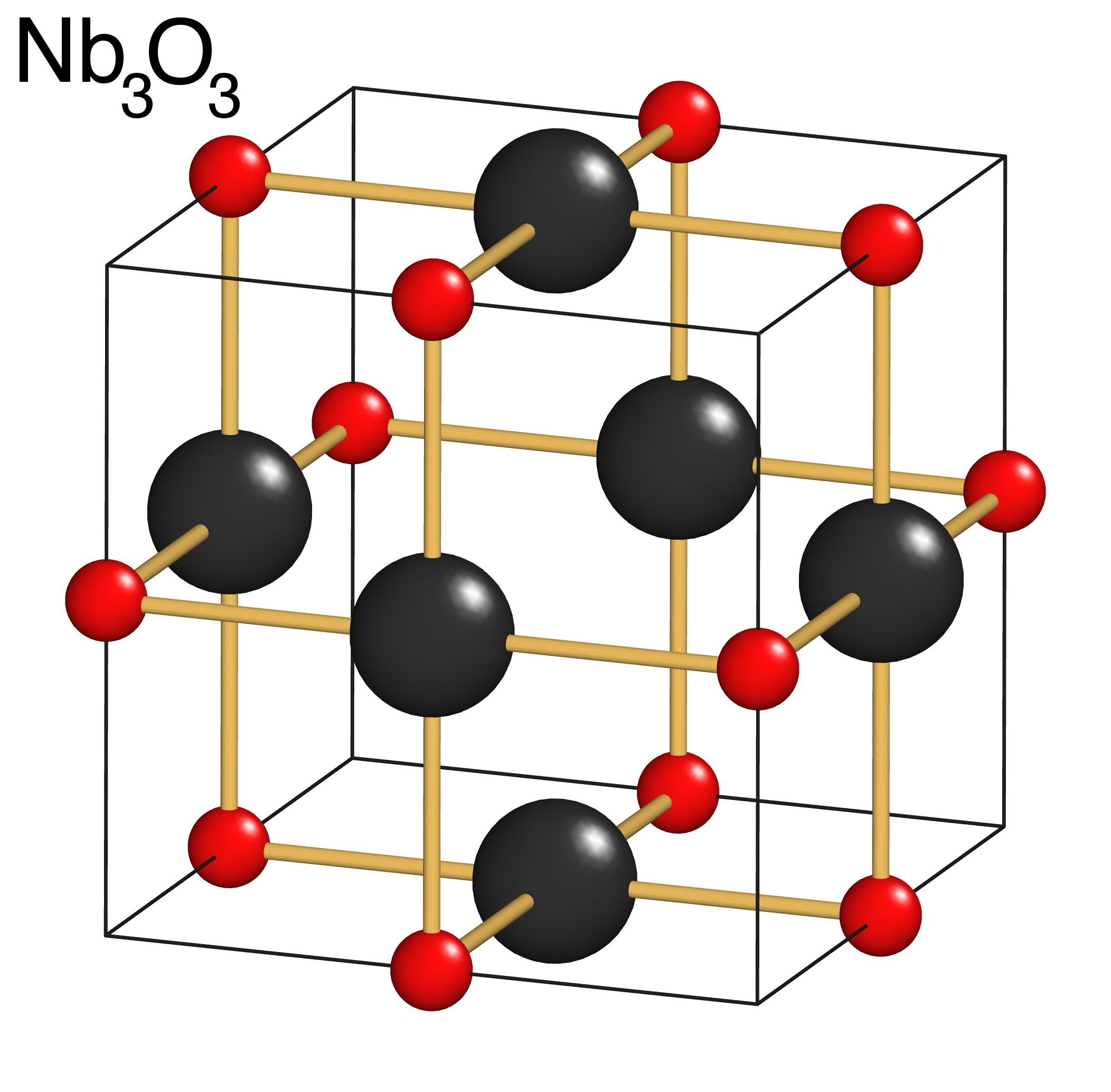}
\includegraphics[clip,width=0.47\columnwidth]{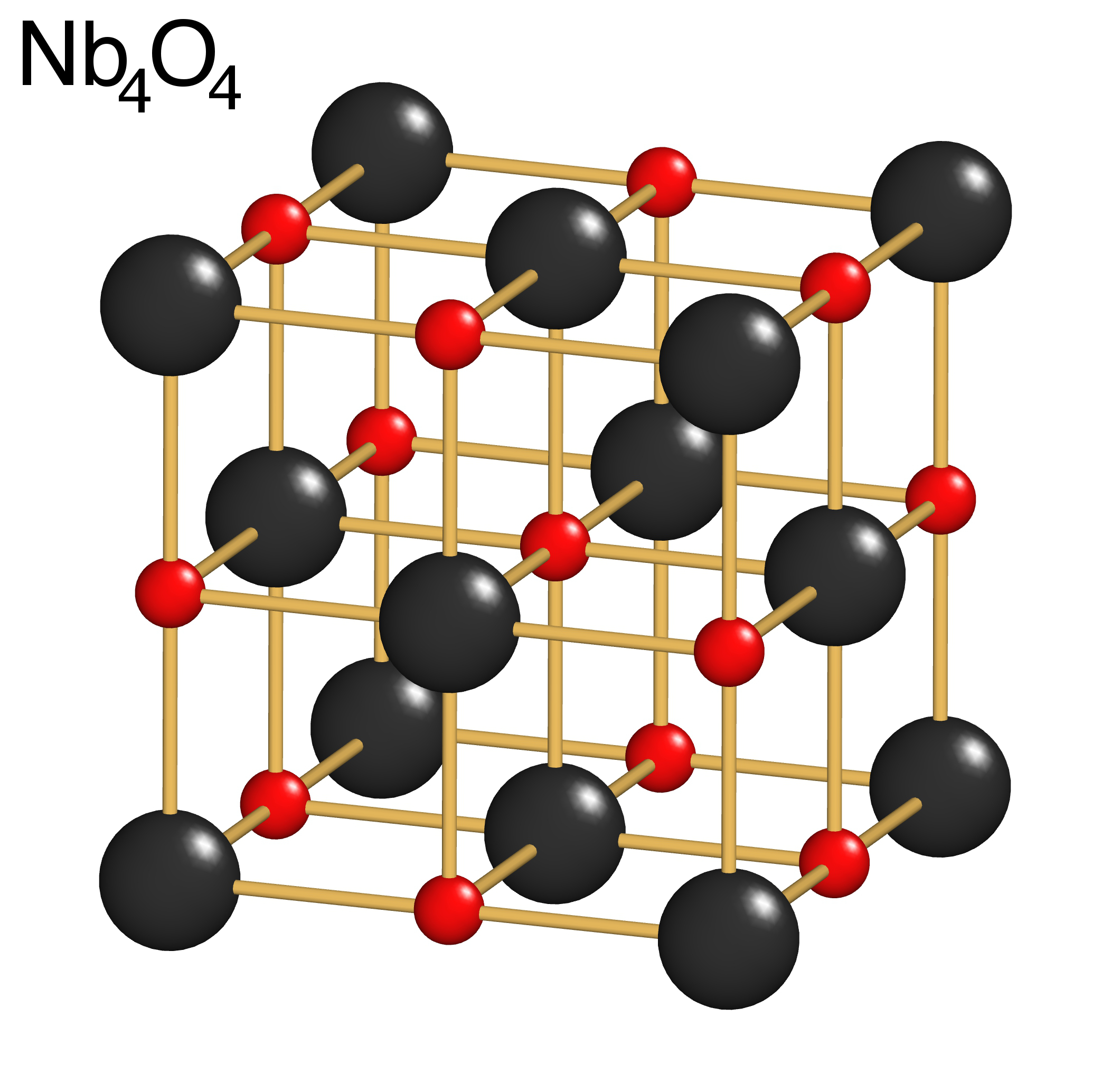}
\caption{Crystal structure of the standard rocksalt lattice (left) and the long-range vacancy-ordered
rocksalt lattice with one quarter of the atoms missing from both sublattices (right). The large and small spheres
depict niobium and oxygen atoms, respectively. Niobium monoxide in the hypothetical rocksalt structure is
labeled as Nb$_{4}$O$_{4}$ and in the real vacancy-ordered structure as Nb$_{3}$O$_{3}$.}
\end{figure}

Angle-integrated photoelectron spectroscopy measurements have been carried out at three facilities:
(1) the Dragon beamline 11A at the National Synchrotron Radiation Research Center (NSRRC) in Taiwan 
with the energy of the soft x-rays set to $h\nu$ = 700\,eV, (2) at the Max-Planck-Institute photoemission 
facility in Dresden having a monochromatized Al-$K_{\alpha}$ $h\nu$ = 1486.6\,eV x-ray source, and 
(3) at the Max-Planck-NSRRC hard x-ray photoelectron spectrosopy (HAXPES) end-station \cite{Weinen_2015} 
at the Taiwan undulator beamline BL12XU of SPring-8 in Japan with the photon energy set to $h\nu$ = 6.5\,keV. 
The photoemission facilities in Taiwan (1) and Japan (3) were equipped with MB Scientific A-1 electron
energy analyzers, and the one in Dresden (2) with a VG Scienta R3000. Angle-resolved photoelectron 
spectroscopy (ARPES) was performed at the (1) NSRRC Dragon beamline 11A with the photon energies varied 
between 90 and 185\,eV in order to cover the fifth Brillouin zone of the niobium monoxide. 
Single crystals of niobium monoxide were grown by the traveling-solvent floating zone method. 
The samples were {\it{ex-situ}} aligned using a Laue camera and cleaved {\it{in-situ}} under ultra-high 
vacuum conditions. The [001] crystal direction is set at normal emission to the electron analyzer lens opening.

The electronic structure calculations were performed using WIEN2k, an augmented plane wave plus local 
orbitals program \cite{wien2k}. Two kinds of parametrization of the exchange-correlation potential were 
employed: the Perdew, Burke and Ernzerhof (PBE) parametrization within the generalized gradient 
approximation (GGA) \cite{Perdew_1996} and a screened hybrid functional for all the electrons \cite{Heyd_2003,*Heyd_2006,Heyd_2004,Tran_2011}.  
The screened hybrid functional ($E_{xc}^{\mathrm{hybrid}}$) 
was constructed such that a part ($\alpha$) of the semi local PBE-GGA exchange ($E_{x}^{\mathrm{GGA}}$) 
was replaced by the short-range part of the Hartree-Fock exchange ($E_{x}^{\mathrm{HF}}$) according to,
\begin{equation}
E_{xc}^{\mathrm{hybrid}} = \alpha E_{x}^{\mathrm{HF}} + (1 - \alpha)E_{x}^{\mathrm{GGA}} + E_{c}^{\mathrm{GGA}}
\end{equation}
where $E_{c}^{\mathrm{GGA}}$ is the correlation energy. We varied $\alpha$ from 0 to 0.3, and the
best fit was found for $\alpha$ = 0.14 as will be shown below. The Brillouin zone was sampled by a well converged 
mesh of 5000 $k$-points in the full zone. The experimental lattice constant $a$ = 4.21\,\AA\,has been used
throughout \cite{Bowman_1966}.

\begin{figure}[t]
\includegraphics[clip,width=\columnwidth]{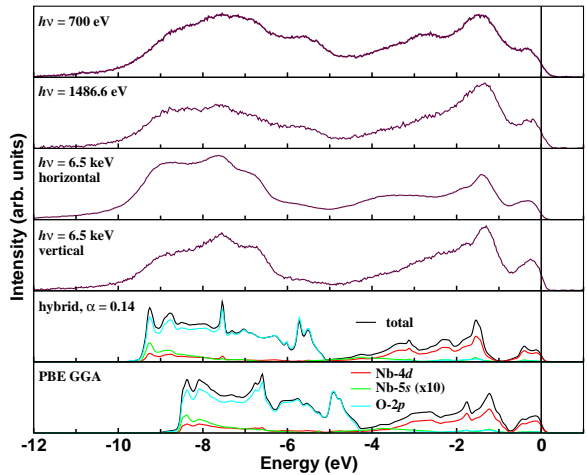}
\caption{Valence band spectra of niobium monoxide taken with photon energies 
$h\nu$ = 700\,eV, 1486.6\,eV and 6.5\,keV together with the total and orbital projected partial
density of states calculated using the hybrid functional and PBE-GGA exchange-correlation 
functional.}
\end{figure}

Figure\,2 displays the valence band (VB) photoemission spectra of niobium monoxide taken with $h\nu$ = 700\,eV, 
1486.6\,eV and 6.5\,keV photon energies. The 700\,eV spectrum was taken at room temperature with an overall
energy resolution of 0.7\,eV, the 1486.6\,eV spectrum at room temperature with 0.4\,eV resolution, and the 6.5\,keV
spectra at 80\,K with 0.17\,eV resolution. All spectra are normalized to their integrated intensities after the 
subtraction of the standard integral background to account for inelastic scattering processes \cite{Hufner_2003}. 
All the spectra show a clear cutoff at zero energy ($E_{\mathrm{F}}$, Fermi level), consistent with the system 
being a good metal \cite{Roberson_1969,Hulm_1972,Honig_1973,Okaz_1975}.
All spectra are comparable to one another in terms of the total band width and peak positions, affirming their 
intrinsic nature. The intensities of the spectral features vary with photon energy, reflecting the photon energy 
dependence of the photo-ionization cross-sections of the atomic orbitals contributing to the VB: the Nb-$4d$, 
O-$2p$ and also the Nb-$5s$ \cite{Yeh_1985,Trzhaskovskaya1,Trzhaskovskaya2,Trzhaskovskaya3}. 

The 6.5\,keV spectra were taken with two different geometries. In the so-called horizontal geometry, the 
electron energy analyzer was mounted horizontally and parallel to the electrical field vector of the photon beam,
while in the vertical geometry, it is perpendicular to the electrical field vector and the Poynting vector of the
beam \cite{Weinen_2015}. The spectral intensities depend strongly on the polarization of the light and is given 
by the so-called $\beta$-asymmetry parameter of the photo-ionization cross-sections of the various atomic shells 
involved \cite{Yeh_1985,Trzhaskovskaya1,Trzhaskovskaya2,Trzhaskovskaya3}. In particular, it has been shown 
experimentally \cite{Weinen_2015}, that the $s$ contribution to the spectra is substantially reduced in the vertical 
geometry. The differences in the 6.5\,keV spectra for the two geometries thus gives an indication for the 
energy distribution of the Nb-$5s$ states. 

VB photoemission spectra have been reported previously for very low energy incident photons with
$h\nu$ = 30, 68 and 83.8\,eV \cite{Honig_1976,Erbudak_1978}. Considering the enormous energy difference of 
the incident photons with the present experiments, the peaks around -1.5\,eV and -7\,eV energies, along with a 
shoulder around -5.5\,eV are qualitatively similar with those earlier reports \cite{Honig_1976,Erbudak_1978}.
Yet, we are now able to discern additional features due to the improved energy resolution and larger variation
of the photo-ionization cross-sections, which help us to present a more precise analysis of the electronic structure
of niobium monoxide as discussed below.

We begin by comparing the experimental VB spectra with the calculated density of states (DOS) for 
Nb$_{3}$O$_{3}$ using the PBE-GGA exchange-correlation functional (Fig.\,2, bottom panel). 
To facilitate an easy comparison, the calculated DOS was multiplied with the Fermi function. 
On the first glance, the general features of the experimental data seem to be reproduced by these 
calculations, with the spectral weight closest to the Fermi level (0 to -4\,eV) originating mainly from 
Nb-4$d$ states, while the spectral weight at deeper energies (-5 to -10\,eV) are an admixture 
of Nb-4$d$ and O-2$p$ states. On closer inspection, certain discrepancies emerge: first and foremost, 
the width of the measured spectra ($\sim$\,10\,eV) is larger than the calculations ($\sim$\,8.5\,eV), 
thereby reducing the separation between the centers of the Nb-4$d$ and O-2$p$ derived spectral 
weight. As a consequence, the shoulder around -5.5\,eV in the experiment is seen at -4.8\,eV in 
the calculated DOS. Additionally, the Nb-4$d$ derived peak at -1.5\,eV energy in the 
experimental spectra is shifted slightly closer to the Fermi level in the calculations. Such discrepancies 
are reminiscent of spurious {\it self-interaction} effects not sufficiently accounted for in conventional DFT.

An approach to improve the orbital energies is to make use of hybrid functionals where an admixture 
of the exact Hartree-Fock exchange to conventional DFT functionals is incorporated \cite{Heyd_2003,*Heyd_2006,Heyd_2004}. 
We therefore have carried out such calculations by also varying the mixing parameter $\alpha$ (see Eq.\,1)
from 0 to 0.30, applied not only to the Nb-4$d$ states, but to all electrons in the system. The optimal mixing 
parameter for Nb$_{3}$O$_{3}$ turned out to be 0.14, smaller than the standard value of 0.25 generally 
used for semiconducting or insulating 3$d$ TM-oxides, and perhaps not inconsistent considering the fact 
that niobium monoxide is a $4d$ system and a good metal too. The resulting Nb-4$d$, O-2$p$ and Nb-5$s$ 
partial-DOS are plotted in Fig.\,2. The width of the VB using hybrid functional is more extended than 
that of PBE-GGA and agrees better with the experimental spectra. The O-2$p$ derived shoulder around 
-5.5\,eV, the Nb-4$d$ peak around -1.5\,eV and the van-Hove-like peak around -7.5\,eV are precisely 
reproduced in our calculations. The hybrid-functional approach thus provides a highly accurate 
description of the valence band spectrum. 

\begin{figure}[t]
\includegraphics[clip,width=\columnwidth]{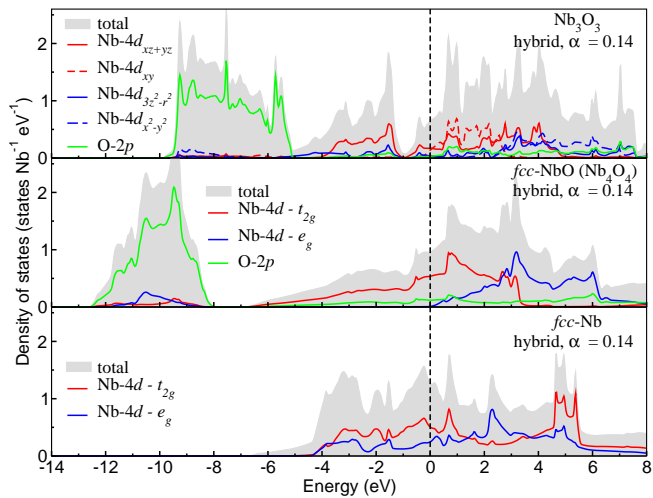}
\caption{Total and projected density of states (DOS) for Nb$_3$O$_3$ (top panel) and the 
hypothetical $fcc$-Nb$_4$O$_4$ (middle panel) using the screened hybrid functional with 
$\alpha$ = 0.14. The bottom panel shows the DOS for Nb metal in the $fcc$ crystal structure. A lattice constant 
of $a$ = 4.21\,\AA\,has been used for all three calculations.}
\end{figure}

It should be noted here that the presence of defect-free $fcc$-Nb$_4$O$_4$ was previously already deemed 
hypothetical based on the finding that the Nb$_{3}$O$_{3}$ structure is more stable than the Nb$_{4}$O$_{4}$ 
by about 1\,eV per NbO  unit \cite{Schwarz_1984,Schulz_1993}. Also, the comparison of the DOS to the then
available spectroscopic data supported this notion \cite{Honig_1976,Wimmer_1982}. To confirm that such is still 
the case even when considering 
{\it self-interaction} corrections, we performed hybrid functional calculations for Nb$_4$O$_4$ as well, 
with the same lattice constant $a$ = 4.21\,\AA\,as Nb$_{3}$O$_{3}$. The resulting total and projected DOS are presented 
in Fig.\,3: indeed the DOS of Nb$_4$O$_4$  (middle panel) are very different from that of Nb$_3$O$_3$ or
to our experimental VB spectra (Fig.\,2), thus confirming the hypothetical nature of $fcc$-Nb$_4$O$_4$.

One immediate striking difference between the DOS of Nb$_{4}$O$_{4}$ and that of Nb$_{3}$O$_{3}$ 
is the much larger energy spread of the occupied DOS of the Nb$_{4}$O$_{4}$, i.e. extending to -12.5\,eV 
energy, while the DOS of Nb$_{3}$O$_{3}$ ends already at -10\,eV. The fact that the O $2p$ derived states in 
Nb$_{4}$O$_{4}$ extend to such a deep energy and become separated with a gap from the Nb $5d$ derived
states reflects the higher Madelung potential in the denser Nb$_{4}$O$_{4}$ structure. Therefore, the smaller 
Madelung energy of Nb$_{3}$O$_{3}$ must be (over)compensated by another electronic mechanism for 
the structure to be more stable. 

The top and middle panels of Fig.\,3 show a break down of the Nb-4$d$ orbitals contributing to the VB. One can clearly 
observe that the $e_g$ bands of the Nb$_{4}$O$_{4}$ structure are essentially above the Fermi level. In
Nb$_{3}$O$_{3}$, the situation is quite different: part of the $e_g$ band, namely the $4d_{3z^2-r^2}$, is now 
extending well below the Fermi level. This constitutes a gain in the formation energy and its origin has been discussed 
in the past in terms of a local cluster \cite{Schafer_1964, Andersen_1984,Burdett_1984}. In Nb$_4$O$_4$, each Nb is 
octahedrally coordinated by six O ions and the hybridization between them lead to a splitting of the Nb-4$d$ levels into 
$t_{2g}$ and $e_g$ states. Since the $\pi$-bonding for the $t_{2g}$ is significantly smaller than the $\sigma$-bonding for 
the $e_g$, the lowest Nb-$4d$ states are derived from the $t_{2g}$ states. With the Nb having the formal 2+ valence
and thus the $4d^3$ configuration, the occupied states are then made of only the $t_{2g}$ while the $e_g$ remain
essentially unoccupied, see middle panel of Fig.\,3. In Nb$_{3}$O$_{3}$, the Nb is locally square planar coordinated 
by four O ions. The lack of "apical" oxygens in this coordination makes that there is no $\sigma$-bond for the 
4$d_{3z^{2}-r^{2}}$ orbital so that this state does not get pushed up in energy by the O-$2p$. Together with the 
large inter-$4d$ hybridization, the 4$d_{3z^{2}-r^{2}}$ can develop bands, part of which is low enough in energy 
to become occupied \cite{Schafer_1964, Andersen_1984, Burdett_1984,Burdett_1993,Schulz_1993,Burdett_1995}, 
see top panel of Fig.\,3. Obviously, with the 4$d_{3z^{2}-r^{2}}$ becoming partially occupied, the $t_{2g}$ should 
move up somewhat to conserve the number of electrons, but apparently this does not cost too much energy, 
so that in the end there is a net gain of 1\,eV per NbO  unit \cite{Schwarz_1984,Schulz_1993} for Nb$_{3}$O$_{3}$ 
in comparison to Nb$_{4}$O$_{4}$.

It is important to note that the inter-$4d$ hybridization is quite large. This can be illustrated by calculating
the band structure of $fcc$ Nb metal with the same lattice constant as the hypothetical $fcc$-Nb$_{4}$O$_{4}$.
The results are presented in the bottom panel of Fig.\,3. One can observe that both the Nb-$4d$ $t_{2g}$ and $e_g$
bands have width of roughly 9\,eV. Such a width implies that correlation effects will not have a chance to 
stabilize a magnetic or insulating states, and that a non-magnetic metallic solution for the ground state of
Nb$_{4}$O$_{4}$ and also Nb$_{3}$O$_{3}$ will be preferred. It is also precisely this large inter-$4d$ band width
which allows for the partial filling of the 4$d_{3z^{2}-r^{2}}$ band once it is not pushed up anymore to 
high energies by the O-$2p$ as the Nb is in a local square planar symmetry in the Nb$_{3}$O$_{3}$ structure.

We would like to point out that the situation in 3$d$ transition metal monoxides is quite different.  We have calculated 
that, for example, the inter-3$d$ band width for $fcc$ V with the VO lattice constant of $a$ = 4.073\,\AA\, is about 
4\,eV and for $fcc$ Ni with the NiO lattice constant of $a$ = 4.176\,\AA\, is about 2\,eV (see Supplemental Material 
\cite{supplement}). With such smaller one-electron band widths, electron correlation effects will manifest with the 
result that magnetic and insulating solutions can (and are) realized, thereby reducing tremendously the effective or
'ARPES' dispersions of the $3d$ derived bands. Thus for $3d$ oxides one may very well need quite different energy 
considerations to explain the formation of defects which then also will have a more localized nature. A band structure 
approach alone may not be adequate. 

\begin{figure}[t]
\includegraphics[clip,width=\columnwidth]{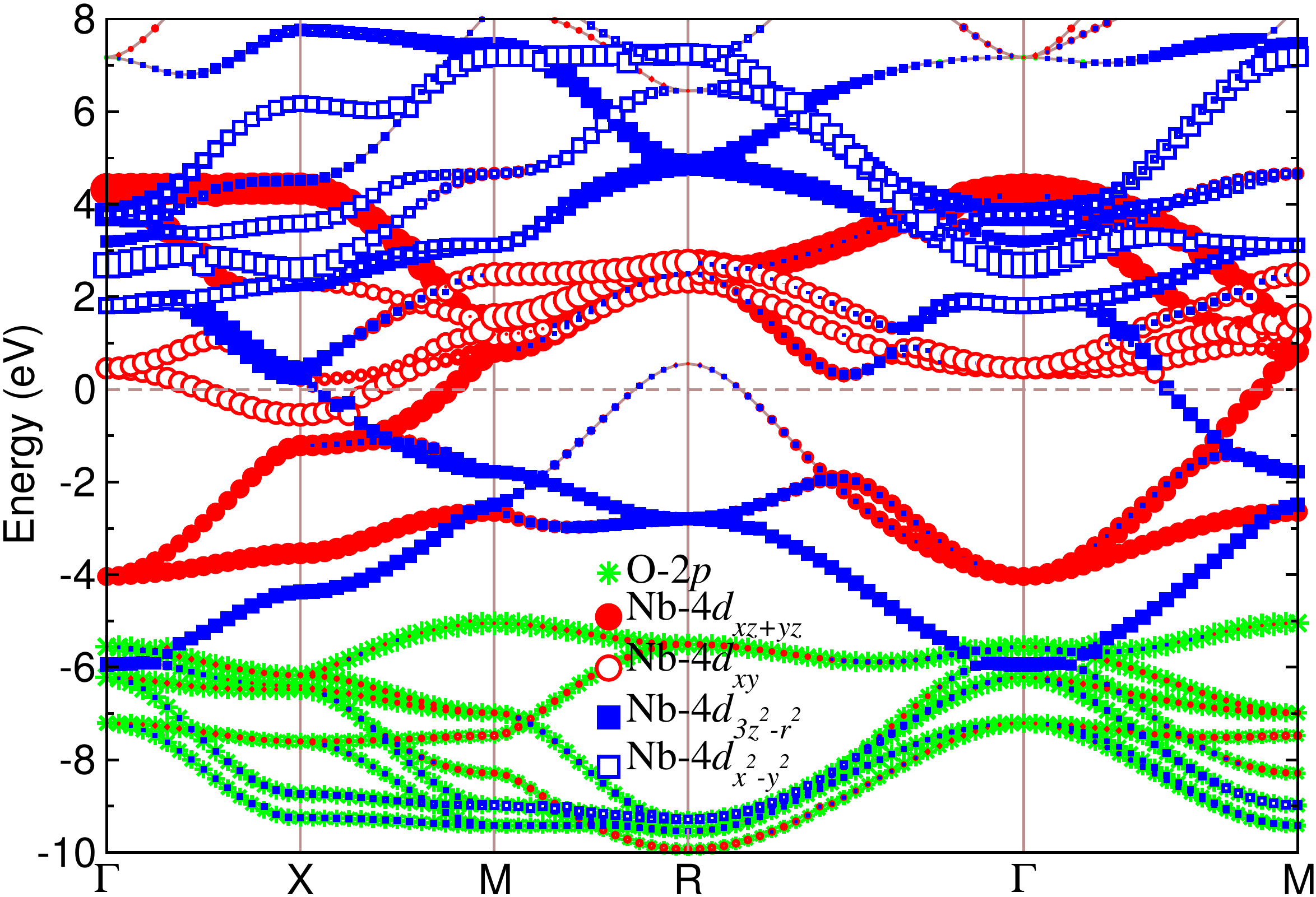}
\caption{Calculated band structure including orbital character for Nb$_3$O$_3$ 
using the screened hybrid functional with $\alpha$ = 0.14.}
\end{figure}

With the inter-4$d$ band width in niobium monoxide being very large, the interaction between the vacancies will 
also be long ranged, so that ordering of the vacancies can be readily expected and consequently, 'vacancy' bands are 
formed \cite{Wimmer_1982,Schulz_1993}. To illustrate this, we display in Fig.\,4 the $k$-dependent band structure 
where we have put labels for the orbital character of the various bands. These results are very similar to 
earlier studies \cite{Wimmer_1982,Kubo_1986,Aoki_1990,Schulz_1993} but with the difference that we used hybrid
functionals with $\alpha$ = 0.14 in order to have the best agreement with the experiment (Fig.\,2).
One can clearly see that below the Fermi level there are heavily dispersing bands with a predominantly 
Nb-4$d_{3z^{2}-r^{2}}$ character. These are the 'vacancy' bands of niobium monoxide 
\cite{Wimmer_1982,Schulz_1993}.

\begin{figure}[t]
\includegraphics[clip,width=\columnwidth]{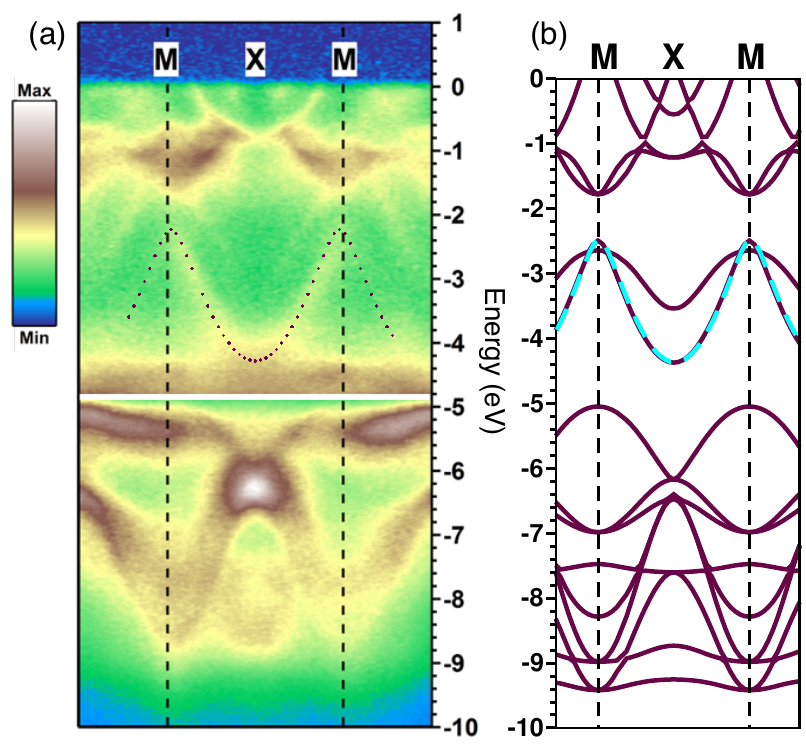}
\caption{(a) Experimental angle-resolved photoelectron spectroscopy (ARPES) images along 
$M-X-M$  with $h\nu$ = 168\,eV.  (b) Calculated band structure along $M-X-M$ using the screened hybrid functional with $\alpha$ = 0.14. The 'vacancy-band' is highlighted using dotted/dashed lines in both panels.}
\end{figure}

Our next task is to show experimentally that these vacancy bands indeed exist, and by that, to justify the
band structure approach for the understanding of the vacancy formation and ordering of the vacancies
in niobium monoxide. Fig.\,5(a) depicts the experimental intensity image of the ARPES spectrum of niobium
monoxide along the $M-X-M$ path measured at $h\nu$ = 168\,eV (see Supplemental Material for further experimental details \cite{supplement}). The dispersing features in the image
can be grouped into three energy sections: (i) the electron and hole pockets spanning from $E_{\mathrm{F}}$ 
to -1.4\,eV; (ii) one medium and one strongly dispersive band spanning -2\,eV to -4.5\,eV; (iii) a dense set 
of oxygen bands below -5\,eV. Fig.\,5(b) displays the calculated band structure along the same $M-X-M$ path.
We can observe good general agreement between the measured and calculated band structure. In all the 
three energy sections mentioned above the band energy positions and dispersions are well reproduced.
Of utmost relevance to our work is the middle section where we can observe a dispersive band which we
have highlighted using dotted/dashed lines in Fig.\,5(a) and (b). This is the Nb-4$d_{3z^{2}-r^{2}}$ 'vacancy' band
as predicted by the band structure calculations. 

To summarize, we have performed angle-integrated and angle-resolved photoelectron spectroscopy 
measurements on niobium monoxide to investigate its electronic structure and the relationship with the
vacancy ordered crystal structure. We have established that band theory provides a good approach
and can describe the valence band features accurately by taking into account also corrections for 
{\it self-interaction} effects. The large inter-Nb-$4d$ band width plays an important role in the stabilization
and ordering of the vacancies and this is clearly demonstrated by the identification of the so-called
'vacancy' band in the experimental angle-resolved spectra. 

We thank A. A. Tsirlin for fruitful discussions and T. Mende for skillful technical assistance. 
D.K. gratefully acknowledges financial support from the Deutsche Forschungsgemeinschaft through project
FOR 1346.

%\bibliography{Nb3O3}

\begin{thebibliography}{40}%
\makeatletter
\providecommand \@ifxundefined [1]{%
 \@ifx{#1\undefined}
}%
\providecommand \@ifnum [1]{%
 \ifnum #1\expandafter \@firstoftwo
 \else \expandafter \@secondoftwo
 \fi
}%
\providecommand \@ifx [1]{%
 \ifx #1\expandafter \@firstoftwo
 \else \expandafter \@secondoftwo
 \fi
}%
\providecommand \natexlab [1]{#1}%
\providecommand \enquote  [1]{``#1''}%
\providecommand \bibnamefont  [1]{#1}%
\providecommand \bibfnamefont [1]{#1}%
\providecommand \citenamefont [1]{#1}%
\providecommand \href@noop [0]{\@secondoftwo}%
\providecommand \href [0]{\begingroup \@sanitize@url \@href}%
\providecommand \@href[1]{\@@startlink{#1}\@@href}%
\providecommand \@@href[1]{\endgroup#1\@@endlink}%
\providecommand \@sanitize@url [0]{\catcode `\\12\catcode `\$12\catcode
  `\&12\catcode `\#12\catcode `\^12\catcode `\_12\catcode `\%12\relax}%
\providecommand \@@startlink[1]{}%
\providecommand \@@endlink[0]{}%
\providecommand \url  [0]{\begingroup\@sanitize@url \@url }%
\providecommand \@url [1]{\endgroup\@href {#1}{\urlprefix }}%
\providecommand \urlprefix  [0]{URL }%
\providecommand \Eprint [0]{\href }%
\providecommand \doibase [0]{http://dx.doi.org/}%
\providecommand \selectlanguage [0]{\@gobble}%
\providecommand \bibinfo  [0]{\@secondoftwo}%
\providecommand \bibfield  [0]{\@secondoftwo}%
\providecommand \translation [1]{[#1]}%
\providecommand \BibitemOpen [0]{}%
\providecommand \bibitemStop [0]{}%
\providecommand \bibitemNoStop [0]{.\EOS\space}%
\providecommand \EOS [0]{\spacefactor3000\relax}%
\providecommand \BibitemShut  [1]{\csname bibitem#1\endcsname}%
\let\auto@bib@innerbib\@empty
%</preamble>
\bibitem [{\citenamefont {Rose}(1858)}]{Rose_1858}%
  \BibitemOpen
  \bibfield  {author} {\bibinfo {author} {\bibfnamefont {H.}~\bibnamefont
  {Rose}},\ }\bibfield  {title} {\enquote {\bibinfo {title} {Ueber das
  {N}iob},}\ }\href@noop {} {\bibfield  {journal} {\bibinfo  {journal} {Pogg.
  Ann.}\ }\textbf {\bibinfo {volume} {104}},\ \bibinfo {pages} {310} (\bibinfo
  {year} {1858})}\BibitemShut {NoStop}%
\bibitem [{\citenamefont {Brauer}(1941)}]{Brauer_1941}%
  \BibitemOpen
  \bibfield  {author} {\bibinfo {author} {\bibfnamefont {G.}~\bibnamefont
  {Brauer}},\ }\bibfield  {title} {\enquote {\bibinfo {title} {Die {O}xyde des
  {N}iobs},}\ }\href@noop {} {\bibfield  {journal} {\bibinfo  {journal} {Z.
  anorg. allg. Chem.}\ }\textbf {\bibinfo {volume} {248}},\ \bibinfo {pages}
  {1} (\bibinfo {year} {1941})}\BibitemShut {NoStop}%
\bibitem [{\citenamefont {Andersson}\ and\ \citenamefont
  {Magn\'{e}li}(1957)}]{Andersson_1957}%
  \BibitemOpen
  \bibfield  {author} {\bibinfo {author} {\bibfnamefont {G.}~\bibnamefont
  {Andersson}}\ and\ \bibinfo {author} {\bibfnamefont {A.}~\bibnamefont
  {Magn\'{e}li}},\ }\bibfield  {title} {\enquote {\bibinfo {title} {Note on the
  {C}rystal {S}tructure of {N}iobium {M}onoxide},}\ }\href@noop {} {\bibfield
  {journal} {\bibinfo  {journal} {Acta Chem. Scand.}\ }\textbf {\bibinfo
  {volume} {11}},\ \bibinfo {pages} {1065} (\bibinfo {year}
  {1957})}\BibitemShut {NoStop}%
\bibitem [{\citenamefont {Bowman}\ \emph {et~al.}(1966)\citenamefont {Bowman},
  \citenamefont {Wallace}, \citenamefont {Yarnell},\ and\ \citenamefont
  {Wenzel}}]{Bowman_1966}%
  \BibitemOpen
  \bibfield  {author} {\bibinfo {author} {\bibfnamefont {A.~L.}\ \bibnamefont
  {Bowman}}, \bibinfo {author} {\bibfnamefont {T.~C.}\ \bibnamefont {Wallace}},
  \bibinfo {author} {\bibfnamefont {J.~L.}\ \bibnamefont {Yarnell}}, \ and\
  \bibinfo {author} {\bibfnamefont {R.~G.}\ \bibnamefont {Wenzel}},\ }\bibfield
   {title} {\enquote {\bibinfo {title} {The crystal structure of niobium
  monoxide},}\ }\href@noop {} {\bibfield  {journal} {\bibinfo  {journal} {Acta.
  Crystallogr.}\ }\textbf {\bibinfo {volume} {21}},\ \bibinfo {pages} {843}
  (\bibinfo {year} {1966})}\BibitemShut {NoStop}%
\bibitem [{\citenamefont {Kolchin}\ and\ \citenamefont
  {Sumarokova}(1961)}]{Kolchin_1961}%
  \BibitemOpen
  \bibfield  {author} {\bibinfo {author} {\bibfnamefont {O.~P.}\ \bibnamefont
  {Kolchin}}\ and\ \bibinfo {author} {\bibfnamefont {N.~V.}\ \bibnamefont
  {Sumarokova}},\ }\bibfield  {title} {\enquote {\bibinfo {title} {The melting
  point and other properties of the lower oxides of niobium},}\ }\href@noop {}
  {\bibfield  {journal} {\bibinfo  {journal} {Atomnaya Energiya}\ }\textbf
  {\bibinfo {volume} {10}},\ \bibinfo {pages} {168} (\bibinfo {year}
  {1961})}\BibitemShut {NoStop}%
\bibitem [{\citenamefont {Taylor}\ and\ \citenamefont
  {Doyle}(1971{\natexlab{a}})}]{Taylor_1971_1}%
  \BibitemOpen
  \bibfield  {author} {\bibinfo {author} {\bibfnamefont {A.}~\bibnamefont
  {Taylor}}\ and\ \bibinfo {author} {\bibfnamefont {N.~J.}\ \bibnamefont
  {Doyle}},\ }\bibfield  {title} {\enquote {\bibinfo {title} {The thermal
  expansion of titanium, vanadium and niobium monoxides},}\ }\href@noop {}
  {\bibfield  {journal} {\bibinfo  {journal} {J. Appl. Crystallogr.}\ }\textbf
  {\bibinfo {volume} {4}},\ \bibinfo {pages} {103} (\bibinfo {year}
  {1971}{\natexlab{a}})}\BibitemShut {NoStop}%
\bibitem [{\citenamefont {Taylor}\ and\ \citenamefont
  {Doyle}(1971{\natexlab{b}})}]{Taylor_1971_2}%
  \BibitemOpen
  \bibfield  {author} {\bibinfo {author} {\bibfnamefont {A.}~\bibnamefont
  {Taylor}}\ and\ \bibinfo {author} {\bibfnamefont {N.~J.}\ \bibnamefont
  {Doyle}},\ }\bibfield  {title} {\enquote {\bibinfo {title} {Compressibilities
  and {G}r\"{u}neisen constants of the monoxides of titanium, vanadium and
  niobium},}\ }\href@noop {} {\bibfield  {journal} {\bibinfo  {journal} {J.
  Appl. Crystallogr.}\ }\textbf {\bibinfo {volume} {4}},\ \bibinfo {pages}
  {109} (\bibinfo {year} {1971}{\natexlab{b}})}\BibitemShut {NoStop}%
\bibitem [{\citenamefont {Reed}\ \emph {et~al.}(1973)\citenamefont {Reed},
  \citenamefont {Pollard}, \citenamefont {Lonney}, \citenamefont {Loehman},\
  and\ \citenamefont {Honig}}]{Reed_2007}%
  \BibitemOpen
  \bibfield  {author} {\bibinfo {author} {\bibfnamefont {T.~B.}\ \bibnamefont
  {Reed}}, \bibinfo {author} {\bibfnamefont {E.~R.}\ \bibnamefont {Pollard}},
  \bibinfo {author} {\bibfnamefont {L.~E.}\ \bibnamefont {Lonney}}, \bibinfo
  {author} {\bibfnamefont {R.~E.}\ \bibnamefont {Loehman}}, \ and\ \bibinfo
  {author} {\bibfnamefont {J.~M.}\ \bibnamefont {Honig}},\ }\enquote {\bibinfo
  {title} {Niobium monoxide},}\ in\ \href {\doibase 10.1002/9780470132456.ch25}
  {\emph {\bibinfo {booktitle} {Inorganic Syntheses, Volume 14}}},\ \bibinfo
  {editor} {edited by\ \bibinfo {editor} {\bibfnamefont {A.}~\bibnamefont
  {Wold}}\ and\ \bibinfo {editor} {\bibfnamefont {J.~K.}\ \bibnamefont
  {Ruff}}}\ (\bibinfo  {publisher} {John Wiley \& Sons, Inc.},\ \bibinfo {year}
  {1973})\ pp.\ \bibinfo {pages} {131--134}\BibitemShut {NoStop}%
\bibitem [{\citenamefont {Wadsley}(1964)}]{Wadsley_1964}%
  \BibitemOpen
  \bibfield  {author} {\bibinfo {author} {\bibfnamefont {A.~D.}\ \bibnamefont
  {Wadsley}},\ }\enquote {\bibinfo {title} {Inorganic nonstoichiometric
  compounds},}\ in\ \href@noop {} {\emph {\bibinfo {booktitle}
  {Nonstoichiometric Compounds}}},\ \bibinfo {editor} {edited by\ \bibinfo
  {editor} {\bibfnamefont {L.}~\bibnamefont {Mandelcorn}}}\ (\bibinfo
  {publisher} {Academic Press, New York and London},\ \bibinfo {year} {1964})\
  pp.\ \bibinfo {pages} {98--209}\BibitemShut {NoStop}%
\bibitem [{\citenamefont {Sch\"{a}fer}\ and\ \citenamefont {von
  Schnering}(1964)}]{Schafer_1964}%
  \BibitemOpen
  \bibfield  {author} {\bibinfo {author} {\bibfnamefont {H.}~\bibnamefont
  {Sch\"{a}fer}}\ and\ \bibinfo {author} {\bibfnamefont {H.~G.}\ \bibnamefont
  {von Schnering}},\ }\bibfield  {title} {\enquote {\bibinfo {title}
  {{M}etall-{M}etall-{B}indungen bei niederen {H}alogeniden, {O}xyden und
  {O}xydhalogeniden schwerer \"{U}bergangsmetalle {T}hermochemische und
  strukturelle {P}rinzipien},}\ }\href@noop {} {\bibfield  {journal} {\bibinfo
  {journal} {Angew. Chem.}\ }\textbf {\bibinfo {volume} {76}},\ \bibinfo
  {pages} {833} (\bibinfo {year} {1964})}\BibitemShut {NoStop}%
\bibitem [{\citenamefont {Burdett}\ and\ \citenamefont
  {Mitchell}(1995)}]{Burdett_1995}%
  \BibitemOpen
  \bibfield  {author} {\bibinfo {author} {\bibfnamefont {J.~K.}\ \bibnamefont
  {Burdett}}\ and\ \bibinfo {author} {\bibfnamefont {J.~F.}\ \bibnamefont
  {Mitchell}},\ }\bibfield  {title} {\enquote {\bibinfo {title}
  {Nonstoichiometry in early transition metal compounds with the rocksalt
  structure},}\ }\href@noop {} {\bibfield  {journal} {\bibinfo  {journal}
  {Prog. Solid St. Chem.}\ }\textbf {\bibinfo {volume} {23}},\ \bibinfo {pages}
  {131} (\bibinfo {year} {1995})}\BibitemShut {NoStop}%
\bibitem [{\citenamefont {Andersen}\ and\ \citenamefont
  {Satpathy}(1984)}]{Andersen_1984}%
  \BibitemOpen
  \bibfield  {author} {\bibinfo {author} {\bibfnamefont {O.~K.}\ \bibnamefont
  {Andersen}}\ and\ \bibinfo {author} {\bibfnamefont {S.}~\bibnamefont
  {Satpathy}},\ }\enquote {\bibinfo {title} {Calculation of the electronic
  bandstructure for the 3$d$-monoxides and the vacancy compound
  {Nb$_{3}$O$_{3}$}},}\ in\ \href@noop {} {\emph {\bibinfo {booktitle} {Basic
  properties of binary oxides}}},\ \bibinfo {editor} {edited by\ \bibinfo
  {editor} {\bibfnamefont {A.}~\bibnamefont {Dominguez~Rodriguez}}, \bibinfo
  {editor} {\bibfnamefont {J.}~\bibnamefont {Casting}}, \ and\ \bibinfo
  {editor} {\bibfnamefont {R.}~\bibnamefont {Marquez}}}\ (\bibinfo  {publisher}
  {Servicio de Publicationes de la Universidad de Sevilla (Serie Ciencias),
  Sevilla},\ \bibinfo {year} {1984})\ pp.\ \bibinfo {pages}
  {21--42}\BibitemShut {NoStop}%
\bibitem [{\citenamefont {Chevrel}\ and\ \citenamefont
  {Sergent}(1986)}]{Chevrel_1986}%
  \BibitemOpen
  \bibfield  {author} {\bibinfo {author} {\bibfnamefont {R.}~\bibnamefont
  {Chevrel}}\ and\ \bibinfo {author} {\bibfnamefont {M.}~\bibnamefont
  {Sergent}},\ }\enquote {\bibinfo {title} {From three-dimensional to
  one-dimensional cluster {Mo$_6$} chalcogenides},}\ in\ \href {\doibase
  10.1007/978-94-009-4528-9_8} {\emph {\bibinfo {booktitle} {Crystal Chemistry
  and Properties of Materials with Quasi-One-Dimensional Structures: A Chemical
  and Physical Synthetic Approach}}},\ \bibinfo {editor} {edited by\ \bibinfo
  {editor} {\bibfnamefont {Jean}\ \bibnamefont {Rouxel}}}\ (\bibinfo
  {publisher} {Springer Netherlands},\ \bibinfo {address} {Dordrecht},\
  \bibinfo {year} {1986})\ pp.\ \bibinfo {pages} {315--373}\BibitemShut
  {NoStop}%
\bibitem [{\citenamefont {Wimmer}\ \emph {et~al.}(1982)\citenamefont {Wimmer},
  \citenamefont {Schwarz}, \citenamefont {Podloucky}, \citenamefont {Herzig},\
  and\ \citenamefont {Neckel}}]{Wimmer_1982}%
  \BibitemOpen
  \bibfield  {author} {\bibinfo {author} {\bibfnamefont {E.}~\bibnamefont
  {Wimmer}}, \bibinfo {author} {\bibfnamefont {K.}~\bibnamefont {Schwarz}},
  \bibinfo {author} {\bibfnamefont {R.}~\bibnamefont {Podloucky}}, \bibinfo
  {author} {\bibfnamefont {P.}~\bibnamefont {Herzig}}, \ and\ \bibinfo {author}
  {\bibfnamefont {A.}~\bibnamefont {Neckel}},\ }\bibfield  {title} {\enquote
  {\bibinfo {title} {The effect of vacancies on the electronic structure of
  {NbO}},}\ }\href@noop {} {\bibfield  {journal} {\bibinfo  {journal} {J. Phys.
  Chem. Solids}\ }\textbf {\bibinfo {volume} {43}},\ \bibinfo {pages} {439}
  (\bibinfo {year} {1982})}\BibitemShut {NoStop}%
\bibitem [{\citenamefont {Schwarz}(1984)}]{Schwarz_1984}%
  \BibitemOpen
  \bibfield  {author} {\bibinfo {author} {\bibfnamefont {K.~H.}\ \bibnamefont
  {Schwarz}},\ }\enquote {\bibinfo {title} {Band structure and
  non-stoichiometry of metallic oxides},}\ in\ \href@noop {} {\emph {\bibinfo
  {booktitle} {Basic properties of binary oxides}}},\ \bibinfo {editor} {edited
  by\ \bibinfo {editor} {\bibfnamefont {A.}~\bibnamefont
  {Dominguez~Rodriguez}}, \bibinfo {editor} {\bibfnamefont {J.}~\bibnamefont
  {Casting}}, \ and\ \bibinfo {editor} {\bibfnamefont {R.}~\bibnamefont
  {Marquez}}}\ (\bibinfo  {publisher} {Servicio de Publicationes de la
  Universidad de Sevilla (Serie Ciencias), Sevilla},\ \bibinfo {year} {1984})\
  pp.\ \bibinfo {pages} {43--56}\BibitemShut {NoStop}%
\bibitem [{\citenamefont {Schulz}\ and\ \citenamefont
  {Wentzcovitch}(1993)}]{Schulz_1993}%
  \BibitemOpen
  \bibfield  {author} {\bibinfo {author} {\bibfnamefont {W.~W.}\ \bibnamefont
  {Schulz}}\ and\ \bibinfo {author} {\bibfnamefont {R.~M.}\ \bibnamefont
  {Wentzcovitch}},\ }\bibfield  {title} {\enquote {\bibinfo {title} {Electronic
  band structure and bonding in {Nb$_{3}$O$_{3}$}},}\ }\href@noop {} {\bibfield
   {journal} {\bibinfo  {journal} {Phys. Rev. B}\ }\textbf {\bibinfo {volume}
  {48}},\ \bibinfo {pages} {16986} (\bibinfo {year} {1993})}\BibitemShut
  {NoStop}%
\bibitem [{\citenamefont {Okaz}\ and\ \citenamefont
  {Keesom}(1975)}]{Okaz_1975}%
  \BibitemOpen
  \bibfield  {author} {\bibinfo {author} {\bibfnamefont {A.~M.}\ \bibnamefont
  {Okaz}}\ and\ \bibinfo {author} {\bibfnamefont {P.~H.}\ \bibnamefont
  {Keesom}},\ }\bibfield  {title} {\enquote {\bibinfo {title} {Specific heat
  and magnetization of the superconducting monoxides: {NbO} and {TiO}},}\
  }\href@noop {} {\bibfield  {journal} {\bibinfo  {journal} {Phys. Rev. B}\
  }\textbf {\bibinfo {volume} {12}},\ \bibinfo {pages} {4917} (\bibinfo {year}
  {1975})}\BibitemShut {NoStop}%
\bibitem [{\citenamefont {Honig}\ \emph {et~al.}(1976)\citenamefont {Honig},
  \citenamefont {Sinha}, \citenamefont {Wahnsiedler},\ and\ \citenamefont
  {Kuwamoto}}]{Honig_1976}%
  \BibitemOpen
  \bibfield  {author} {\bibinfo {author} {\bibfnamefont {J.~M.}\ \bibnamefont
  {Honig}}, \bibinfo {author} {\bibfnamefont {A.~P.~B.}\ \bibnamefont {Sinha}},
  \bibinfo {author} {\bibfnamefont {W.~E.}\ \bibnamefont {Wahnsiedler}}, \ and\
  \bibinfo {author} {\bibfnamefont {H.}~\bibnamefont {Kuwamoto}},\ }\bibfield
  {title} {\enquote {\bibinfo {title} {Studies of the {B}and {S}tructure of
  {NbO} by {X}-{R}ay {P}hotoelectron {S}pectroscopy},}\ }\href@noop {}
  {\bibfield  {journal} {\bibinfo  {journal} {Phys. Stat. Sol.}\ }\textbf
  {\bibinfo {volume} {B73}},\ \bibinfo {pages} {651} (\bibinfo {year}
  {1976})}\BibitemShut {NoStop}%
\bibitem [{\citenamefont {Erbudak}\ \emph {et~al.}(1978)\citenamefont
  {Erbudak}, \citenamefont {Gubanov},\ and\ \citenamefont
  {Kurmaev}}]{Erbudak_1978}%
  \BibitemOpen
  \bibfield  {author} {\bibinfo {author} {\bibfnamefont {M.}~\bibnamefont
  {Erbudak}}, \bibinfo {author} {\bibfnamefont {V.~A.}\ \bibnamefont
  {Gubanov}}, \ and\ \bibinfo {author} {\bibfnamefont {E.~Z.}\ \bibnamefont
  {Kurmaev}},\ }\bibfield  {title} {\enquote {\bibinfo {title} {The electronic
  structure of {NbO}: Theory and experiment},}\ }\href@noop {} {\bibfield
  {journal} {\bibinfo  {journal} {J. Phys. Chem. Solids}\ }\textbf {\bibinfo
  {volume} {39}},\ \bibinfo {pages} {1157} (\bibinfo {year}
  {1978})}\BibitemShut {NoStop}%
\bibitem [{\citenamefont {Schulz}\ \emph {et~al.}(1992)\citenamefont {Schulz},
  \citenamefont {Forro}, \citenamefont {Kendziora}, \citenamefont
  {Wentzcovitch}, \citenamefont {Mandrus}, \citenamefont {Mihaly},\ and\
  \citenamefont {Allen}}]{Schulz_1992}%
  \BibitemOpen
  \bibfield  {author} {\bibinfo {author} {\bibfnamefont {W.~W.}\ \bibnamefont
  {Schulz}}, \bibinfo {author} {\bibfnamefont {L.}~\bibnamefont {Forro}},
  \bibinfo {author} {\bibfnamefont {C.}~\bibnamefont {Kendziora}}, \bibinfo
  {author} {\bibfnamefont {R.}~\bibnamefont {Wentzcovitch}}, \bibinfo {author}
  {\bibfnamefont {D.}~\bibnamefont {Mandrus}}, \bibinfo {author} {\bibfnamefont
  {L.}~\bibnamefont {Mihaly}}, \ and\ \bibinfo {author} {\bibfnamefont {P.~B.}\
  \bibnamefont {Allen}},\ }\bibfield  {title} {\enquote {\bibinfo {title} {Band
  structure and electronic transport properties of the superconductor {NbO}},}\
  }\href@noop {} {\bibfield  {journal} {\bibinfo  {journal} {Phys. Rev. B}\
  }\textbf {\bibinfo {volume} {46}},\ \bibinfo {pages} {46} (\bibinfo {year}
  {1992})}\BibitemShut {NoStop}%
\bibitem [{\citenamefont {Weinen}\ \emph {et~al.}(2015)\citenamefont {Weinen},
  \citenamefont {Koethe}, \citenamefont {Chang}, \citenamefont {Agrestini},
  \citenamefont {Kasinathan}, \citenamefont {Liao}, \citenamefont {Fujiwara},
  \citenamefont {Sch\"{u}{\ss}ler-Langeheine}, \citenamefont {Strigari},
  \citenamefont {Haupricht}, \citenamefont {Panaccione}, \citenamefont {Offi},
  \citenamefont {Monaco}, \citenamefont {H.}, \citenamefont {Tsuei},\ and\
  \citenamefont {Tjeng}}]{Weinen_2015}%
  \BibitemOpen
  \bibfield  {author} {\bibinfo {author} {\bibfnamefont {J.}~\bibnamefont
  {Weinen}}, \bibinfo {author} {\bibfnamefont {T.~C.}\ \bibnamefont {Koethe}},
  \bibinfo {author} {\bibfnamefont {C.~F.}\ \bibnamefont {Chang}}, \bibinfo
  {author} {\bibfnamefont {S.}~\bibnamefont {Agrestini}}, \bibinfo {author}
  {\bibfnamefont {D.}~\bibnamefont {Kasinathan}}, \bibinfo {author}
  {\bibfnamefont {Y.~F.}\ \bibnamefont {Liao}}, \bibinfo {author}
  {\bibfnamefont {H.}~\bibnamefont {Fujiwara}}, \bibinfo {author}
  {\bibfnamefont {C.}~\bibnamefont {Sch\"{u}{\ss}ler-Langeheine}}, \bibinfo
  {author} {\bibfnamefont {F.}~\bibnamefont {Strigari}}, \bibinfo {author}
  {\bibfnamefont {T.}~\bibnamefont {Haupricht}}, \bibinfo {author}
  {\bibfnamefont {G.}~\bibnamefont {Panaccione}}, \bibinfo {author}
  {\bibfnamefont {F.}~\bibnamefont {Offi}}, \bibinfo {author} {\bibfnamefont
  {G.}~\bibnamefont {Monaco}}, \bibinfo {author} {\bibfnamefont {Huotari.}\
  \bibnamefont {H.}}, \bibinfo {author} {\bibfnamefont {K.-D.}\ \bibnamefont
  {Tsuei}}, \ and\ \bibinfo {author} {\bibfnamefont {L.~H.}\ \bibnamefont
  {Tjeng}},\ }\bibfield  {title} {\enquote {\bibinfo {title} {Polarization
  dependent hard x-ray photoemission experiments for solids: Efficiency and
  limits for unraveling the orbital character of the valence band},}\
  }\href@noop {} {\bibfield  {journal} {\bibinfo  {journal} {J. Elec.
  Spectroscopy and Related Phenomena}\ }\textbf {\bibinfo {volume} {198}},\
  \bibinfo {pages} {6} (\bibinfo {year} {2015})}\BibitemShut {NoStop}%
\bibitem [{\citenamefont {Blaha}\ \emph {et~al.}(2001)\citenamefont {Blaha},
  \citenamefont {Schwarz}, \citenamefont {Madsen}, \citenamefont {Kvasnicka},\
  and\ \citenamefont {Luitz}}]{wien2k}%
  \BibitemOpen
  \bibfield  {author} {\bibinfo {author} {\bibfnamefont {P.}~\bibnamefont
  {Blaha}}, \bibinfo {author} {\bibfnamefont {K.}~\bibnamefont {Schwarz}},
  \bibinfo {author} {\bibfnamefont {G.~K.~H.}\ \bibnamefont {Madsen}}, \bibinfo
  {author} {\bibfnamefont {D.}~\bibnamefont {Kvasnicka}}, \ and\ \bibinfo
  {author} {\bibfnamefont {J.}~\bibnamefont {Luitz}},\ }\href@noop {} {\emph
  {\bibinfo {title} {{WIEN2k}, {A}n {A}ugmented {P}lane {W}ave + {L}ocal
  {O}rbitals {P}rogram for {C}alculating {C}rystal {P}roperties}}}\ (\bibinfo
  {publisher} {{K}arlheinz Schwarz, Techn. Universit\"{a}t Wien, Austria},\
  \bibinfo {year} {2001})\BibitemShut {NoStop}%
\bibitem [{\citenamefont {Perdew}\ \emph {et~al.}(1996)\citenamefont {Perdew},
  \citenamefont {Burke},\ and\ \citenamefont {Ernzerhof}}]{Perdew_1996}%
  \BibitemOpen
  \bibfield  {author} {\bibinfo {author} {\bibfnamefont {J.~P.}\ \bibnamefont
  {Perdew}}, \bibinfo {author} {\bibfnamefont {K.}~\bibnamefont {Burke}}, \
  and\ \bibinfo {author} {\bibfnamefont {M.}~\bibnamefont {Ernzerhof}},\
  }\bibfield  {title} {\enquote {\bibinfo {title} {Generalized gradient
  approximation made simple},}\ }\href@noop {} {\bibfield  {journal} {\bibinfo
  {journal} {Phys. Rev. Lett.}\ }\textbf {\bibinfo {volume} {77}},\ \bibinfo
  {pages} {3865} (\bibinfo {year} {1996})}\BibitemShut {NoStop}%
\bibitem [{\citenamefont {Heyd}\ \emph {et~al.}(2003)\citenamefont {Heyd},
  \citenamefont {Scuseria},\ and\ \citenamefont {Ernzerhof}}]{Heyd_2003}%
  \BibitemOpen
  \bibfield  {author} {\bibinfo {author} {\bibfnamefont {J.}~\bibnamefont
  {Heyd}}, \bibinfo {author} {\bibfnamefont {G.~E.}\ \bibnamefont {Scuseria}},
  \ and\ \bibinfo {author} {\bibfnamefont {M.}~\bibnamefont {Ernzerhof}},\
  }\bibfield  {title} {\enquote {\bibinfo {title} {Hybrid functionals based on
  a screened {C}oulomb potential},}\ }\href@noop {} {\bibfield  {journal}
  {\bibinfo  {journal} {J. Chem. Phys.}\ }\textbf {\bibinfo {volume} {118}},\
  \bibinfo {pages} {8207} (\bibinfo {year} {2003})}\BibitemShut {NoStop}%
\bibitem [{\citenamefont {Heyd}\ \emph {et~al.}(2006)\citenamefont {Heyd},
  \citenamefont {Scuseria},\ and\ \citenamefont {Ernzerhof}}]{Heyd_2006}%
  \BibitemOpen
  \bibfield  {author} {\bibinfo {author} {\bibfnamefont {J.}~\bibnamefont
  {Heyd}}, \bibinfo {author} {\bibfnamefont {G.~E.}\ \bibnamefont {Scuseria}},
  \ and\ \bibinfo {author} {\bibfnamefont {M.}~\bibnamefont {Ernzerhof}},\
  }\href@noop {} {\bibfield  {journal} {\bibinfo  {journal} {J. Chem. Phys.}\
  }\textbf {\bibinfo {volume} {124}},\ \bibinfo {pages} {219906} (\bibinfo
  {year} {2006})}\BibitemShut {NoStop}%
\bibitem [{\citenamefont {Heyd}\ and\ \citenamefont
  {Scuseria}(2004)}]{Heyd_2004}%
  \BibitemOpen
  \bibfield  {author} {\bibinfo {author} {\bibfnamefont {J.}~\bibnamefont
  {Heyd}}\ and\ \bibinfo {author} {\bibfnamefont {G.~E.}\ \bibnamefont
  {Scuseria}},\ }\bibfield  {title} {\enquote {\bibinfo {title} {Efficient
  hybrid density functional calculations in solids: assessment of the
  {H}eyd-{S}cuseria-{E}rnzerhof screened {C}oulomb hybrid functional},}\
  }\href@noop {} {\bibfield  {journal} {\bibinfo  {journal} {J. Chem. Phys.}\
  }\textbf {\bibinfo {volume} {121}},\ \bibinfo {pages} {1187} (\bibinfo {year}
  {2004})}\BibitemShut {NoStop}%
\bibitem [{\citenamefont {Tran}\ and\ \citenamefont {Blaha}(2011)}]{Tran_2011}%
  \BibitemOpen
  \bibfield  {author} {\bibinfo {author} {\bibfnamefont {F.}~\bibnamefont
  {Tran}}\ and\ \bibinfo {author} {\bibfnamefont {P.}~\bibnamefont {Blaha}},\
  }\bibfield  {title} {\enquote {\bibinfo {title} {Implementation of screened
  hybrid functionals based on the {Y}ukawa potential within the {LAPW} basis
  set},}\ }\href@noop {} {\bibfield  {journal} {\bibinfo  {journal} {Phys. Rev.
  B}\ }\textbf {\bibinfo {volume} {83}},\ \bibinfo {pages} {235118} (\bibinfo
  {year} {2011})}\BibitemShut {NoStop}%
\bibitem [{\citenamefont {H{\"u}fner}(2003)}]{Hufner_2003}%
  \BibitemOpen
  \bibfield  {author} {\bibinfo {author} {\bibfnamefont {Stefan}\ \bibnamefont
  {H{\"u}fner}},\ }\enquote {\bibinfo {title} {Continuous satellites and
  plasmon satellites: {XPS} photoemission in nearly free electron systems},}\
  in\ \href {\doibase 10.1007/978-3-662-09280-4_4} {\emph {\bibinfo {booktitle}
  {Photoelectron Spectroscopy: Principles and Applications}}}\ (\bibinfo
  {publisher} {Springer Berlin Heidelberg},\ \bibinfo {year} {2003})\ pp.\
  \bibinfo {pages} {173--209}\BibitemShut {NoStop}%
\bibitem [{\citenamefont {Roberson}\ and\ \citenamefont
  {Rapp}(1969)}]{Roberson_1969}%
  \BibitemOpen
  \bibfield  {author} {\bibinfo {author} {\bibfnamefont {J.~A.}\ \bibnamefont
  {Roberson}}\ and\ \bibinfo {author} {\bibfnamefont {R.~A.}\ \bibnamefont
  {Rapp}},\ }\bibfield  {title} {\enquote {\bibinfo {title} {Electrical
  properties of {NbO} and {NbO$_{2}$}},}\ }\href@noop {} {\bibfield  {journal}
  {\bibinfo  {journal} {J. Phys. Chem. Solids}\ }\textbf {\bibinfo {volume}
  {30}},\ \bibinfo {pages} {1119} (\bibinfo {year} {1969})}\BibitemShut
  {NoStop}%
\bibitem [{\citenamefont {Hulm}\ \emph {et~al.}(1972)\citenamefont {Hulm},
  \citenamefont {Jones}, \citenamefont {Hein},\ and\ \citenamefont
  {Gibson}}]{Hulm_1972}%
  \BibitemOpen
  \bibfield  {author} {\bibinfo {author} {\bibfnamefont {J.~K.}\ \bibnamefont
  {Hulm}}, \bibinfo {author} {\bibfnamefont {C.~K.}\ \bibnamefont {Jones}},
  \bibinfo {author} {\bibfnamefont {R.~A.}\ \bibnamefont {Hein}}, \ and\
  \bibinfo {author} {\bibfnamefont {J.~W.}\ \bibnamefont {Gibson}},\ }\bibfield
   {title} {\enquote {\bibinfo {title} {Superconductivity in the {TiO} and
  {NbO} systems},}\ }\href@noop {} {\bibfield  {journal} {\bibinfo  {journal}
  {J. Low Temp. Phys.}\ }\textbf {\bibinfo {volume} {7}},\ \bibinfo {pages}
  {291} (\bibinfo {year} {1972})}\BibitemShut {NoStop}%
\bibitem [{\citenamefont {Honig}\ \emph {et~al.}(1973)\citenamefont {Honig},
  \citenamefont {Wahnsiedler},\ and\ \citenamefont {Eklund}}]{Honig_1973}%
  \BibitemOpen
  \bibfield  {author} {\bibinfo {author} {\bibfnamefont {J.~M.}\ \bibnamefont
  {Honig}}, \bibinfo {author} {\bibfnamefont {W.~E.}\ \bibnamefont
  {Wahnsiedler}}, \ and\ \bibinfo {author} {\bibfnamefont {P.~C.}\ \bibnamefont
  {Eklund}},\ }\bibfield  {title} {\enquote {\bibinfo {title} {Electrical
  properties of {NbO} in high magnetic fields},}\ }\href@noop {} {\bibfield
  {journal} {\bibinfo  {journal} {J. Solid State Chem.}\ }\textbf {\bibinfo
  {volume} {6}},\ \bibinfo {pages} {203} (\bibinfo {year} {1973})}\BibitemShut
  {NoStop}%
\bibitem [{\citenamefont {Yeh}\ and\ \citenamefont {Lindau}(1985)}]{Yeh_1985}%
  \BibitemOpen
  \bibfield  {author} {\bibinfo {author} {\bibfnamefont {J.}~\bibnamefont
  {Yeh}}\ and\ \bibinfo {author} {\bibfnamefont {I.}~\bibnamefont {Lindau}},\
  }\bibfield  {title} {\enquote {\bibinfo {title} {Atomic subshell
  photoionization cross sections and asymmetry parameters: 1 $\leq$ {Z} $\leq$
  103},}\ }\href@noop {} {\bibfield  {journal} {\bibinfo  {journal} {Atomic
  Data and Nuclear Tables}\ }\textbf {\bibinfo {volume} {32}},\ \bibinfo
  {pages} {1} (\bibinfo {year} {1985})}\BibitemShut {NoStop}%
\bibitem [{\citenamefont {Trzhaskovskaya}\ \emph {et~al.}(2001)\citenamefont
  {Trzhaskovskaya}, \citenamefont {Nefedov},\ and\ \citenamefont
  {Yarzhemsky}}]{Trzhaskovskaya1}%
  \BibitemOpen
  \bibfield  {author} {\bibinfo {author} {\bibfnamefont {M.~B.}\ \bibnamefont
  {Trzhaskovskaya}}, \bibinfo {author} {\bibfnamefont {V.~I.}\ \bibnamefont
  {Nefedov}}, \ and\ \bibinfo {author} {\bibfnamefont {V.~G.}\ \bibnamefont
  {Yarzhemsky}},\ }\bibfield  {title} {\enquote {\bibinfo {title}
  {Photoelectron angular distribution parameters for elements {Z}=1 to {Z}=54
  in the photoelectron energy range 100-5000 e{V}},}\ }\href@noop {} {\bibfield
   {journal} {\bibinfo  {journal} {Atomic Data and Nuclear Data Tables}\
  }\textbf {\bibinfo {volume} {77}},\ \bibinfo {pages} {97} (\bibinfo {year}
  {2001})}\BibitemShut {NoStop}%
\bibitem [{\citenamefont {Trzhaskovskaya}\ \emph {et~al.}(2002)\citenamefont
  {Trzhaskovskaya}, \citenamefont {Nefedov},\ and\ \citenamefont
  {Yarzhemsky}}]{Trzhaskovskaya2}%
  \BibitemOpen
  \bibfield  {author} {\bibinfo {author} {\bibfnamefont {M.~B.}\ \bibnamefont
  {Trzhaskovskaya}}, \bibinfo {author} {\bibfnamefont {V.~I.}\ \bibnamefont
  {Nefedov}}, \ and\ \bibinfo {author} {\bibfnamefont {V.~G.}\ \bibnamefont
  {Yarzhemsky}},\ }\bibfield  {title} {\enquote {\bibinfo {title}
  {Photoelectron angular distribution parameters for elements {Z}=55 to {Z}=100
  in the photoelectron energy range 100-5000 e{V}},}\ }\href@noop {} {\bibfield
   {journal} {\bibinfo  {journal} {Atomic Data and Nuclear Data Tables}\
  }\textbf {\bibinfo {volume} {82}},\ \bibinfo {pages} {257} (\bibinfo {year}
  {2002})}\BibitemShut {NoStop}%
\bibitem [{\citenamefont {Trzhaskovskaya}\ \emph {et~al.}(2006)\citenamefont
  {Trzhaskovskaya}, \citenamefont {Nikulin}, \citenamefont {Nefedov},\ and\
  \citenamefont {Yarzhemsky}}]{Trzhaskovskaya3}%
  \BibitemOpen
  \bibfield  {author} {\bibinfo {author} {\bibfnamefont {M.~B.}\ \bibnamefont
  {Trzhaskovskaya}}, \bibinfo {author} {\bibfnamefont {V.~K.}\ \bibnamefont
  {Nikulin}}, \bibinfo {author} {\bibfnamefont {V.~I.}\ \bibnamefont
  {Nefedov}}, \ and\ \bibinfo {author} {\bibfnamefont {V.~G.}\ \bibnamefont
  {Yarzhemsky}},\ }\bibfield  {title} {\enquote {\bibinfo {title} {Non-dipole
  second order parameters of the photoelectron angular distribution for
  elements {Z} = 1-100 in the photoelectron energy range 1-10 ke{V}},}\
  }\href@noop {} {\bibfield  {journal} {\bibinfo  {journal} {Atomic Data and
  Nuclear Data Tables}\ }\textbf {\bibinfo {volume} {92}},\ \bibinfo {pages}
  {245} (\bibinfo {year} {2006})}\BibitemShut {NoStop}%
\bibitem [{\citenamefont {Burdett}\ and\ \citenamefont
  {Hughbanks}(1984)}]{Burdett_1984}%
  \BibitemOpen
  \bibfield  {author} {\bibinfo {author} {\bibfnamefont {J.~K.}\ \bibnamefont
  {Burdett}}\ and\ \bibinfo {author} {\bibfnamefont {T.}~\bibnamefont
  {Hughbanks}},\ }\bibfield  {title} {\enquote {\bibinfo {title} {{NbO} and
  {TiO}: Structural and {E}lectronic {S}tability of {S}tructures {D}erived from
  {R}ock {S}alt},}\ }\href@noop {} {\bibfield  {journal} {\bibinfo  {journal}
  {J. Am. Chem. Soc.}\ }\textbf {\bibinfo {volume} {106}},\ \bibinfo {pages}
  {3101} (\bibinfo {year} {1984})}\BibitemShut {NoStop}%
\bibitem [{\citenamefont {Burdett}\ and\ \citenamefont
  {Mitchell}(1993)}]{Burdett_1993}%
  \BibitemOpen
  \bibfield  {author} {\bibinfo {author} {\bibfnamefont {J.~K.}\ \bibnamefont
  {Burdett}}\ and\ \bibinfo {author} {\bibfnamefont {J.~F.}\ \bibnamefont
  {Mitchell}},\ }\bibfield  {title} {\enquote {\bibinfo {title} {Pair
  {P}otentials and the {O}rdered {D}efect {S}tructure of {NbO}},}\ }\href@noop
  {} {\bibfield  {journal} {\bibinfo  {journal} {Inorg. Chem.}\ }\textbf
  {\bibinfo {volume} {32}},\ \bibinfo {pages} {5004} (\bibinfo {year}
  {1993})}\BibitemShut {NoStop}%
\bibitem [{sup()}]{supplement}%
  \BibitemOpen
  \href@noop {} {}\bibinfo {note} {See Supplemental Material at
  http://link.aps.org/supplemental/10.1103/PhysRevLett.000.000000 for details
  about the ARPES measurement and additional density of states}\BibitemShut
  {NoStop}%
\bibitem [{\citenamefont {Kubo}\ \emph {et~al.}(1986)\citenamefont {Kubo},
  \citenamefont {Wakoh},\ and\ \citenamefont {Schwarz}}]{Kubo_1986}%
  \BibitemOpen
  \bibfield  {author} {\bibinfo {author} {\bibfnamefont {Y.}~\bibnamefont
  {Kubo}}, \bibinfo {author} {\bibfnamefont {S.}~\bibnamefont {Wakoh}}, \ and\
  \bibinfo {author} {\bibfnamefont {K.}~\bibnamefont {Schwarz}},\ }\bibfield
  {title} {\enquote {\bibinfo {title} {Theoretical {M}omentum {D}istributions
  in {Nb$_{3}$O$_{3}$}},}\ }\href@noop {} {\bibfield  {journal} {\bibinfo
  {journal} {J. Phys. Soc. Japan}\ }\textbf {\bibinfo {volume} {55}},\ \bibinfo
  {pages} {1266} (\bibinfo {year} {1986})}\BibitemShut {NoStop}%
\bibitem [{\citenamefont {Aoki}\ \emph {et~al.}(1990)\citenamefont {Aoki},
  \citenamefont {Asada}, \citenamefont {Hatano}, \citenamefont {Ogawa},
  \citenamefont {Yanase},\ and\ \citenamefont {Koiwa}}]{Aoki_1990}%
  \BibitemOpen
  \bibfield  {author} {\bibinfo {author} {\bibfnamefont {H.}~\bibnamefont
  {Aoki}}, \bibinfo {author} {\bibfnamefont {Y.}~\bibnamefont {Asada}},
  \bibinfo {author} {\bibfnamefont {T.}~\bibnamefont {Hatano}}, \bibinfo
  {author} {\bibfnamefont {K.}~\bibnamefont {Ogawa}}, \bibinfo {author}
  {\bibfnamefont {A.}~\bibnamefont {Yanase}}, \ and\ \bibinfo {author}
  {\bibfnamefont {M.}~\bibnamefont {Koiwa}},\ }\bibfield  {title} {\enquote
  {\bibinfo {title} {Fermi surface of {NbO}},}\ }\href@noop {} {\bibfield
  {journal} {\bibinfo  {journal} {J. Low. Temp. Phys.}\ }\textbf {\bibinfo
  {volume} {81}},\ \bibinfo {pages} {19} (\bibinfo {year} {1990})}\BibitemShut
  {NoStop}%
\end{thebibliography}

%\end{document}

%

\begin{widetext}
\begin{center}
{\large\bf Supplemental Material}
\end{center}
\end{widetext}

%%%%%%%%%% Merge with supplemental materials %%%%%%%%%%
%%%%%%%%%% Prefix a "S" to all equations, figures, tables and reset the counter %%%%%%%%%%
\setcounter{equation}{0}
\setcounter{figure}{0}
\setcounter{table}{0}
\setcounter{page}{1}
\makeatletter
\renewcommand{\theequation}{S\arabic{equation}}
\renewcommand{\thefigure}{S\arabic{figure}}
\renewcommand{\thesection}{S\arabic{section}}
\renewcommand{\bibnumfmt}[1]{[S#1]}
\renewcommand{\citenumfont}[1]{S#1}
%%%%%%%%%% Prefix a "S" to all equations, figures, tables and reset the counter %%%%%%%%%%

\section{(1) Inter-transition-metal $d$ hybridization in rocksalt oxides}       
In order to estimate the inter-transition-metal hybridization strength in transition metal
$4d$ and $3d$ monoxides oxides with the rocksalt crystal structure, we calculate the 
band structure of fcc NbO/VO/NiO and fcc Nb/V/Ni with the same lattice constant as 
NbO/VO/NiO, respectively. See Fig. S1, which also include the $e_g$ and $t_{2g}$ projections.

\begin{figure}[h]
\includegraphics[clip,width=0.9\columnwidth]{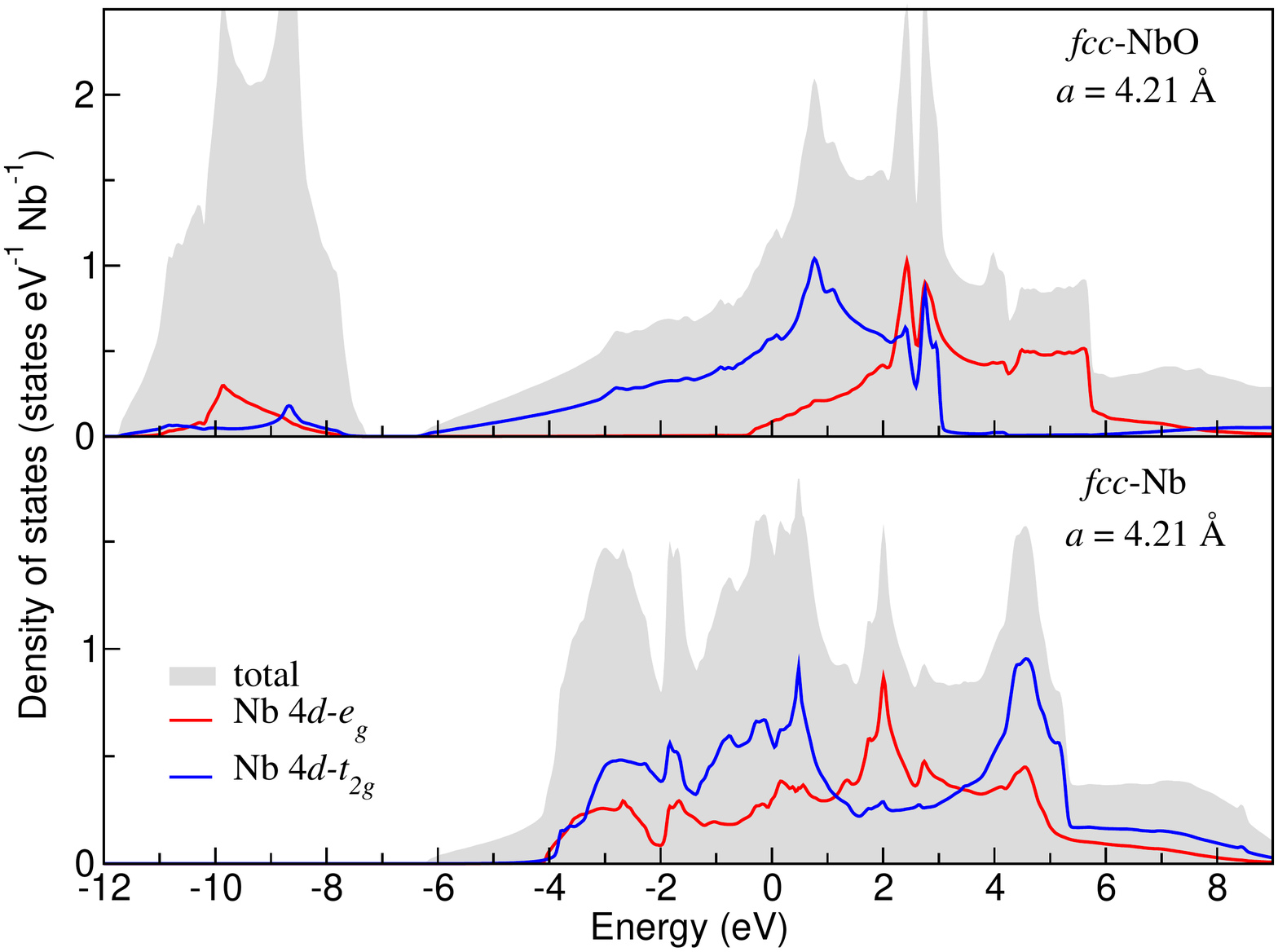}
\includegraphics[clip,width=0.9\columnwidth]{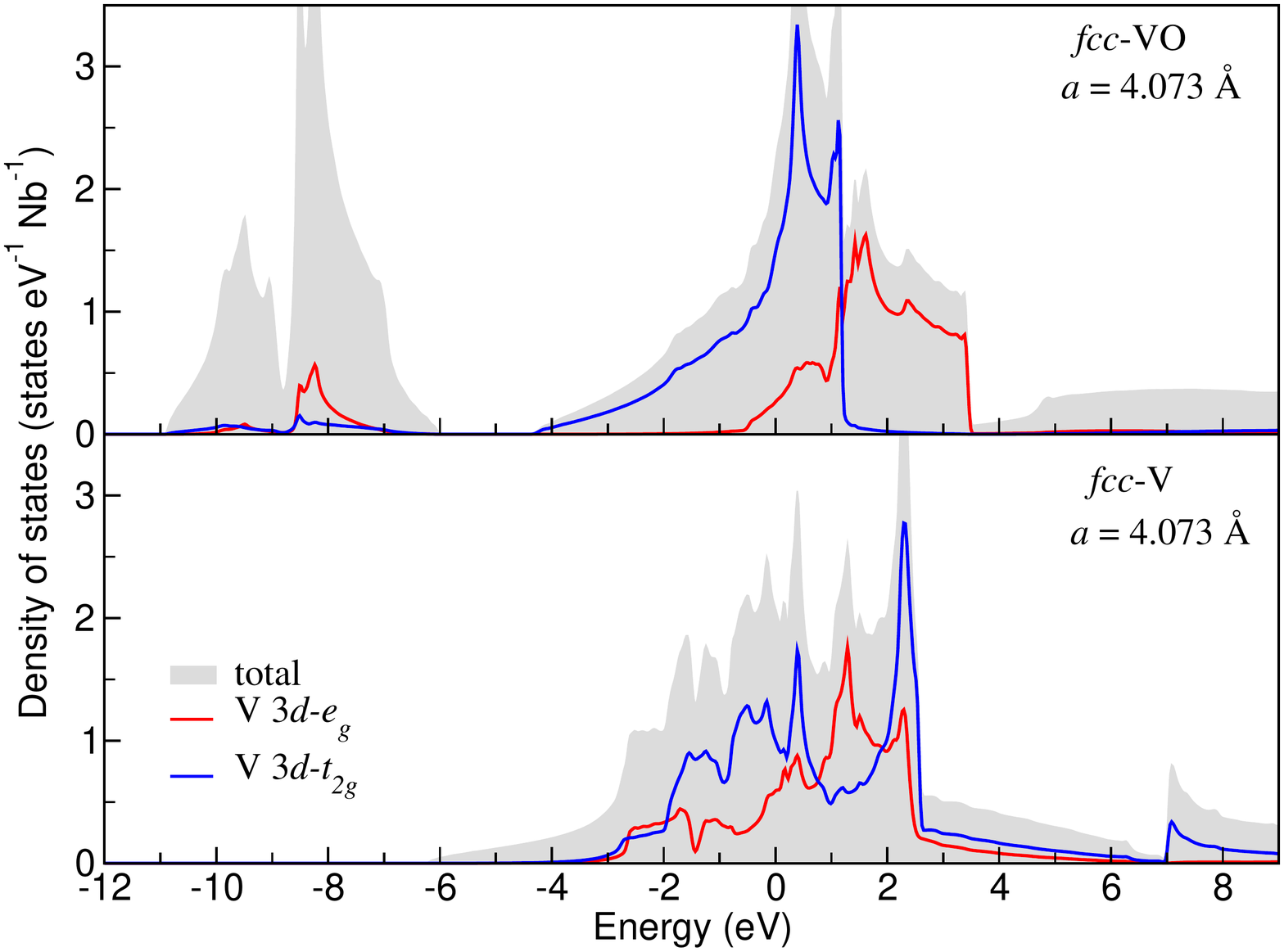}
\includegraphics[clip,width=0.9\columnwidth]{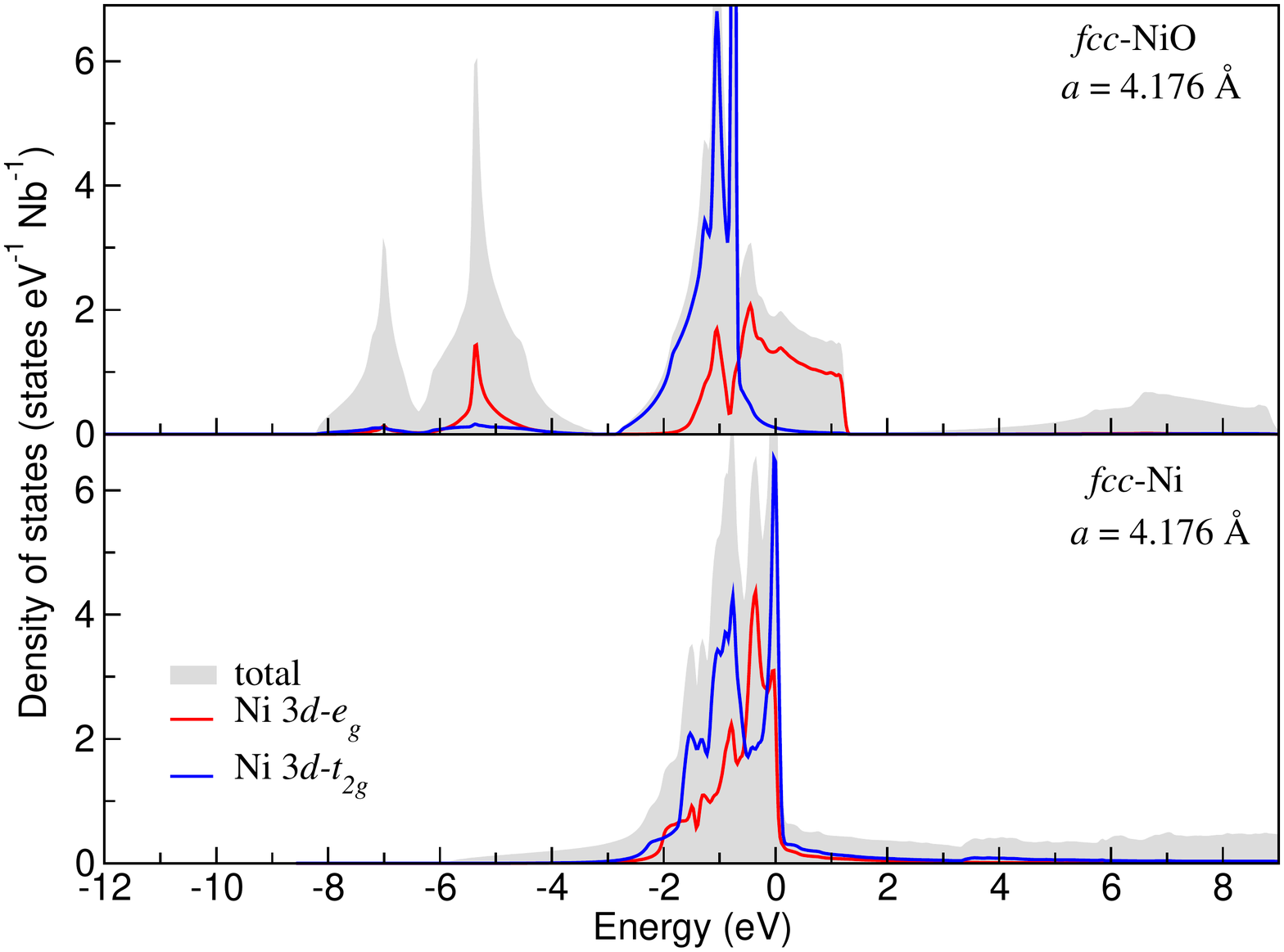}
\caption{\label{dosfcc}Total and $d$-projected density of states for fcc NbO, Nb, VO, V, NiO and Ni. 
The calculations were performed using the WIEN2k code and the Perdew, Burke and Ernzerhof 
(PBE) parametrization within the generalized gradient approximation (GGA).}
\end{figure}

\section{(2) ARPES: momentum space mapping }
The schematics of the momentum space mapping for the niobium monoxide
ARPES experiment is depicted in Fig. S2. The sample is oriented with the [001] 
surface normal directed to the electron energy analyzer. An inner potential 
$V_{0}$ = 9 eV has been used based on the extremal behavior observed in the energy 
dispersions of various Nb and O derived spectral features as function of photon energy.
A photon energy interval of 90 to 185 eV allows for a full coverage of the fifth 
Brillouin zone of niobium monoxide.  

\begin{figure}[t]
\includegraphics[clip,width=\columnwidth]{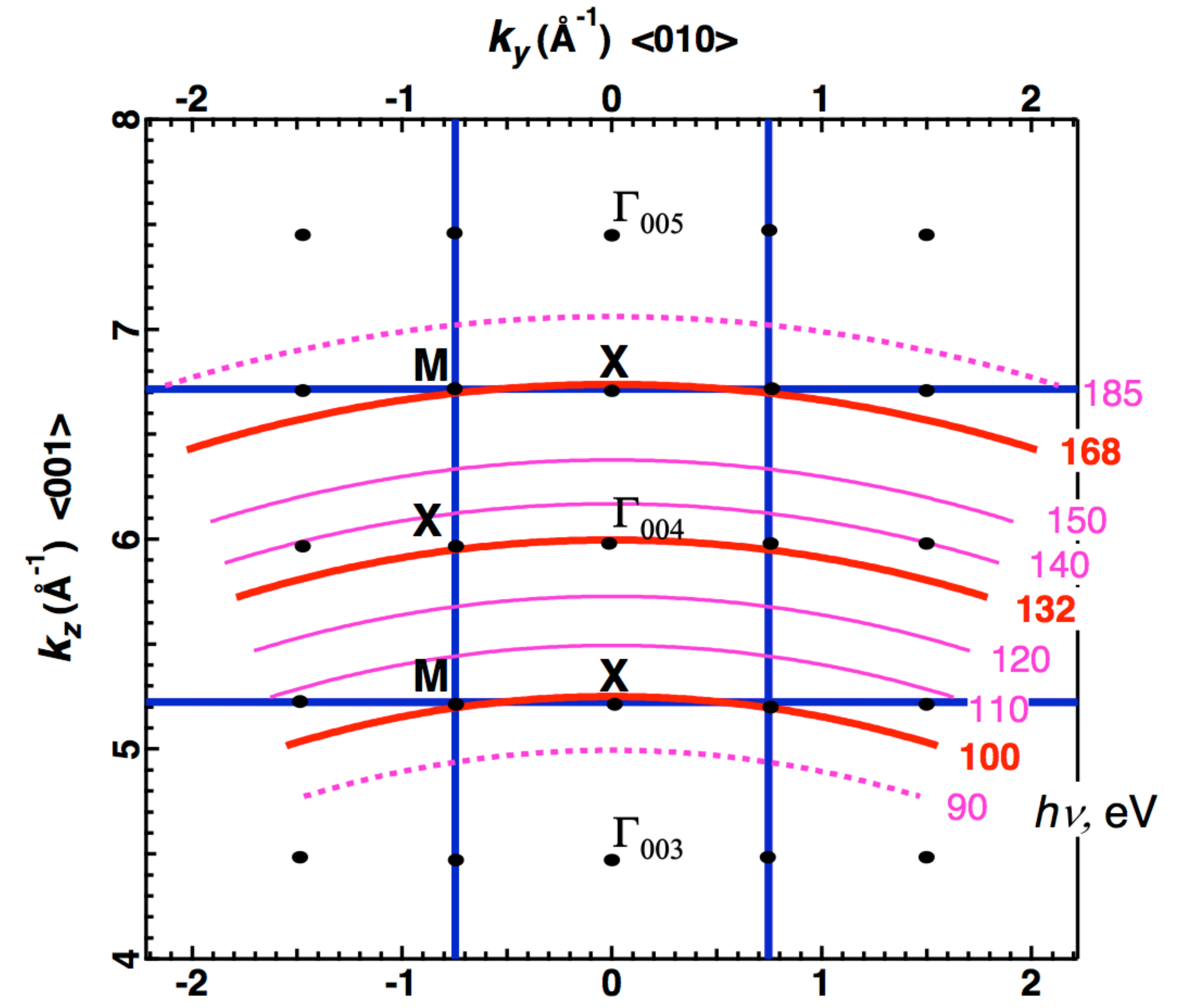}
\caption{\label{bz}Schematics of the experimentally probed cuts in the momentum 
space $k_{y}-k_{z}$ plane with $k_{x}$ = 0. The cuts are for photoelectrons 
with zero binding energy with the corresponding photon energies indicated. An inner 
potential $V_{0}$ = 9 eV has been used. Cuts which pass through the high-symmetry 
points at normal emission are highlighted by the thick red lines. The electron energy 
analyzer has a $\pm$\,18$^\circ$ angular acceptance with a work function of 3.87 eV.
}
\end{figure}

\end{document}